\documentclass{aa}

\usepackage{graphicx}
\usepackage{txfonts}
\graphicspath{{./}{./fig/}}

\begin{document}

\title{Rapid-then-slow migration reproduces mass distribution of TRAPPIST-1 system}

\author{Masahiro Ogihara\inst{1,2,3}
          \and
          Eiichiro Kokubo\inst{3}
          \and
          Ryuunosuke Nakano\inst{4}
          \and
          Takeru K. Suzuki\inst{4}
          }

\institute{Tsung-Dao Lee Institute, Shanghai Jiao Tong University, 520 Shengrong Road, Shanghai 201210, China\\
              \email{ogihara@nagoya-u.jp}
         \and
                        Earth-Life Science Institute, Tokyo Institute of Technology, 2-12-1 Ookayama, Meguro, Tokyo 152-8550, Japan
         \and
             National Astronomical Observatory of Japan, 2-21-1 Osawa, Mitaka, Tokyo 181-8588, Japan
         \and
             School of Arts and Sciences, University of Tokyo, 3-8-1 Komaba, Meguro, Tokyo 153-8902, Japan
             }
             
\date{Received October 4, 2021; accepted January 5, 2022}

\abstract
   {
The TRAPPIST-1 system is an iconic planetary system in various aspects (e.g., habitability, resonant relation, and multiplicity) and hence has attracted considerable attention. The mass distribution of the TRAPPIST-1 planets is characterized by two features: the two inner planets are large, and the masses of the four planets in the outer orbit increase with orbital distance. The origin of these features cannot be explained by previous formation models.
   }
   {
We investigate whether the mass distribution of the TRAPPIST-1 system can be reproduced by a planet formation model using \textit{N}-body simulations.
   }
   {
We used a gas disk evolution model around a low-mass star constructed by considering disk winds and followed the growth and orbital migration from planetary embryos with the isolation mass, which increases with orbital distance.
   }
   {
As a result, we find that from the initial phase, planets in inner orbits undergo rapid orbital migration, and the coalescence growth near the inner disk edge is enhanced. This allows the inner planets to grow larger. Meanwhile, compared with the inner planets, planets in outer orbits migrate more slowly and do not frequently collide with neighboring planets. Therefore, the trend of increasing mass toward the outer orbit, called reversed mass ranking, is maintained.
The final mass distribution approximately agrees with the two features of the mass distribution in the TRAPPIST-1 system.
   }
   {
We discover that the mass distribution in the TRAPPIST-1 system can be reproduced when embryos experience rapid migration and become trapped near the disk inner edge, and then more massive embryos undergo slower migration. This migration transition can be achieved naturally in a disk evolution model with disk winds.
   }

   \maketitle               
                              
\section{Introduction}\label{sec:intro}
TRAPPIST-1 is a late-M dwarf with a mass of $0.089 \,M_\odot$ \citep{2018ApJ...853...30V}.  Seven planets have been discovered in this system. Some of them are thought to exist in the liquid-water habitable zone, and thus have been observed with great interest \citep[e.g.,][]{2016Natur.533..221G,2017Natur.542..456G,2017NatAs...1E.129L}. This system will continue to attract attention from various perspectives. For example, theoretical and observational studies of planetary atmospheres will be extensively conducted in the next decade \citep[e.g.,][]{2016Natur.537...69D,2018NatAs...2..214D,2018AJ....156..178Z,2020ApJ...889...77H,2021arXiv210611444K}.

Based on recent observations, the estimates of the properties of planets in the TRAPPIST-1 system have been updated. \citet{2021PSJ.....2....1A} performed a transit-timing and photometric analysis using extensive transit data, including those obtained by the Spitzer Space Telescope, and refined the estimate of masses and densities of the planets. Table~\ref{tbl:trappist} lists their estimates. Two features in the mass distribution are crucial: the two inner planets b-c have high masses, and the masses of the outer planets d-g increase with orbital distance.

\begin{table}
\caption{Properties of the TRAPPIST-1 planets.}            
\label{tbl:trappist} 
\centering                    
\begin{tabular}{c c c c} 
\hline\hline    
Planet & $a$ ($10^{-2} {\rm \,au}$) & Mass $(M_\oplus)$ & Density $({\rm g\,cm^{-3}})$\\
\hline                
   b & $1.154\pm0.010$ & $1.374\pm0.069$ & $5.425^{+0.265}_{-0.272}$\\      
   c & $1.580\pm0.013$ & $1.308\pm0.056$ & $5.447^{+0.222}_{-0.235}$\\
   d & $2.227\pm0.019$ & $0.388\pm0.012$ & $4.354^{+0.156}_{-0.163}$\\
   e & $2.925\pm0.025$ & $0.692\pm0.022$ & $4.885^{+0.168}_{-0.182}$\\
   f & $3.849\pm0.033$ & $1.039\pm0.031$ & $5.009^{+0.138}_{-0.158}$\\ 
   g & $4.683\pm0.040$ & $1.321\pm0.038$ & $5.042^{+0.136}_{-0.158}$\\ 
   h & $6.189\pm0.053$ & $0.326\pm0.020$ & $4.147^{+0.322}_{-0.302}$\\ 
\hline                     
\end{tabular}
\tablefoot{The second column, $a$, indicates the semimajor axis. Data are obtained from \citet{2021PSJ.....2....1A}.}
\end{table}

The planetary composition can be estimated by modeling the interior structure with the estimates of the planetary mass and radius \citep[e.g.,][]{2018NatAs...2..297U,2018A&A...613A..68G,2018ApJ...865...20D,2021A&A...647A..53A}. The estimates consistently indicate that all seven planets have rocky interiors with a small water mass fraction ($\lesssim 10 {\rm wt\%}$). In particular, the water fraction of the three inner planets is near zero ($\lesssim 10^{-3} {\rm wt\%}$). 
In addition, it has been reported that the densities of the planets can be described with a single mass-radius relation \citep{2021PSJ.....2....1A}.
The orbital configuration of the TRAPPIST-1 planets is also characteristic. The outer planets d-h are in first-order mean-motion resonances (3:2 or 4:3) between neighboring planets. The two neighboring pairs between the inner planets b-d are in near higher-order resonances (8:5 and 5:3).
There are also three-body Laplace resonances \citep{2017NatAs...1E.129L}.

The origin of the TRAPPIST-1 system has been theoretically studied several times. One of the proposed models is that planets formed in outer orbits and migrated inward to the current positions. \citet{2019A&A...631A...7C} performed \textit{N}-body simulations to investigate the model in which planets grow from planetesimals or pebbles in distant orbits ($r \simeq 1-5 {\rm \,au},$ where $r$ is the radial distance from the star) and migrate inward to inner orbits. They showed that the average planetary mass, orbital location, and resonant configuration reasonably agree with those of the TRAPPIST-1 planets. In contrast, the mass distribution shows a trend that inner planets are larger and outer planets are smaller, which does not match the mass distribution for planets d-g in the TRAPPIST-1 system. This mass distribution, called ``mass ranking,'' is a typical result of the formation process in which rapid inward migration is stopped at the inner disk edge \citep{2015A&A...578A..36O,2017MNRAS.470.1750I}. Massive planets grow first and quickly migrate to the disk inner edge. Then, smaller planets successively migrate inward relatively slowly, resulting in capture into mean-motion resonances with the larger inner planets. This process forms the mass ranking.  As for the planetary composition in this migration model, the formed planets contain a large amount of water (up to 50wt\%) because almost all ingredients (planetesimals and pebbles) are made of ice. A similar conclusion was obtained by \citet[][]{2020MNRAS.491.1998M}. \citet{2019A&A...631A...7C} also proposed that to avoid the large water mass fraction, ablation of icy pebbles in the planetary atmosphere and recycling of the polluted atmosphere to the protoplanetary disk are required.

As a different formation model, \citet{2017A&A...604A...1O} proposed that planets grow near the water snowline ($r < 1 {\rm \,au}$), mainly by pebble accretion. In this model, the average planetary mass \citep{2019A&A...627A.149S} and resonant relation \citep{2021ApJ...907...81L,2021arXiv210910984H} could be reproduced. In contrast, the mass was determined by the pebble isolation mass; therefore, the planetary mass was almost uniform in the system. For the planetary composition, although the final water mass fraction is still not small enough ($\simeq 10-20 {\rm wt}\%$), the water fraction can be smaller than that in a large-scale migration model, which suggests that planetary accretion in the close-in region is more suitable for the formation of TRAPPIST-1 system than accretion in distant regions \citep[see also][]{2021arXiv210504596B}. Notably, it is still difficult to achieve zero water mass fraction, as suggested by some recent interior characterization.

The two models described above work well in terms of average mass and resonant relations. However, further investigations are required for the origin of the mass distribution and the composition.
In this study, we investigate the formation of the TRAPPIST-1 system focusing on how to reproduce the mass distribution. To do this, we use the latest disk evolution model. \citet{nakano21} developed a disk evolution model around low-mass stars that considers the effects of disk winds and photoevaporation. Their disk model could explain several observational features of the protoplanetary disk around low-mass stars (e.g., lifetime and spectral energy distribution (SED) slope).

In the disk model, the distribution of the gas surface density differs from the simple power-law distribution used in previous studies. Therefore, the direction and speed of orbital migration are expected to change. It has been shown that the picture of orbital migration changes around solar-type stars when the effects of disk winds are considered \citep[e.g.,][]{2015A&A...584L...1O}. This elucidates several observational features of super-Earths \citep{2018A&A...615A..63O} and of the terrestrial planets in the solar system \citep[][]{2018A&A...612L...5O,2021ApJ...921L...5U}.
As the orbital migration is a crucial factor in determining the final mass distribution of planetary systems, we investigate whether the mass distribution of the TRAPPIST-1 system can be reproduced in a disk evolving with disk winds.

The rest of this paper is structured as follows. 
In Section~\ref{sec:model} we describe the disk model and the numerical method we used in our \textit{N}-body simulation.
In Section~\ref{sec:results} we present the results of the simulation and show how the mass distribution of the TRAPPIST-1 system can be reproduced.
In Section~\ref{sec:discussion} we discuss resonant relations and some caveats in this study.
Finally, in Section~\ref{sec:conc} we conclude this study.

\section{Model}\label{sec:model}
\subsection{Disk evolution model}
\begin{figure}
\centering
\includegraphics[width=\hsize]{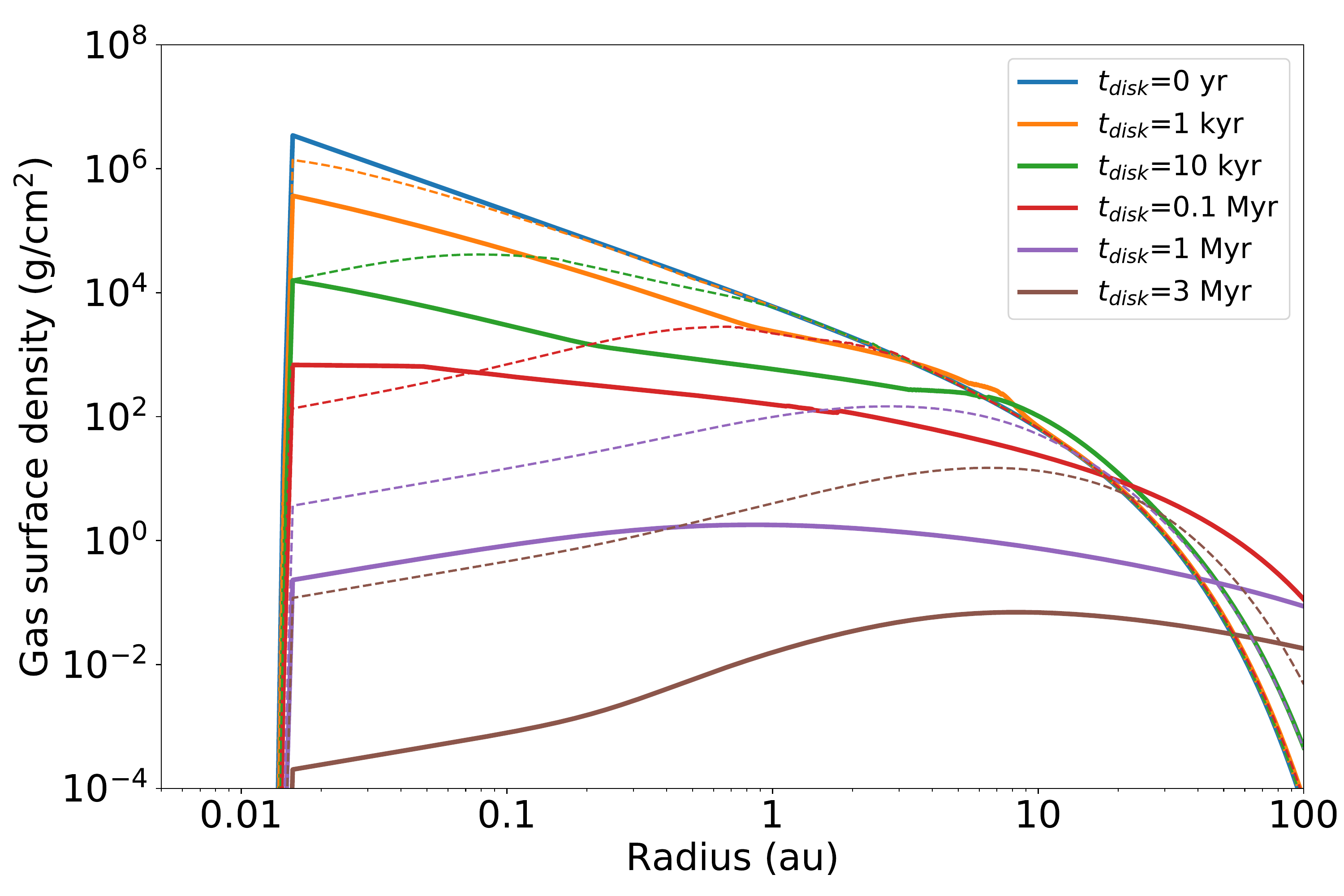}
\caption{Temporal evolution of the gas surface density for our main simulation (solid curves). The dashed curves are for our additional simulation with lower turbulent viscosity. Initially, the distribution has a power-law index of -3/2, but the index changes with time in the inner region due to the influence of disk winds. As in previous studies, the inner disk edge is placed at $r = 0.015 {\rm \,au}$.
The start time of \textit{N}-body simulations is delayed by $t_0$ from $t_{\rm disk}$.
}
\label{fig:r_sigma}
\end{figure}

We used a disk evolution model around a low-mass star that includes effects of disk winds and photoevaporative winds developed by \citet{nakano21}. The temporal evolution of the gas surface density, $\Sigma_{\rm g}$, is given by solving the following equation:
\begin{align}
\frac{\partial \Sigma_{\rm g}}{\partial t} =& \frac{1}{r}\frac{\partial}{\partial r}
\left\{\frac{2}{r\Omega}\left[\frac{\partial}{\partial r}(r^2 \Sigma_{\rm g}
 \alpha_{r\phi}c_{\rm s}^2) + r^2 \alpha_{\phi z}
 \frac{\Sigma_{\rm g}H\Omega^2}{2\sqrt{\pi}} \right]\right\}\nonumber\\
&- C_{\rm w} \frac{ \Sigma_{\rm g}\Omega }{2\sqrt{\pi}} - \dot{\Sigma}_{\rm PE},
\label{eq:sigma}
\end{align}
where $\Omega$, $c_{\rm s}$, and $H$ indicate the Keplerian angular velocity, sound speed, and disk scale height, respectively.
The first term on the right-hand side is the accretion driven by the standard $\alpha$ viscosity \citep{1973A&A....24..337S}, the second term is the wind-driven accretion, the third term is the mass loss due to disk winds, and the last term is the mass loss due to photoevaporation. For our fiducial model, the viscous stress parameter was set to $\alpha_{r,\phi}=8 \times 10^{-3}$, unless otherwise stated. For prescriptions of parameters for disk winds (i.e., $\alpha_{\phi z}$ and $C_{\rm w}$), see \citet{2016A&A...596A..74S}. See also \citet{2020MNRAS.492.3849K} and \citet{nakano21} for details of  mass loss due to photoevaporation, $\dot{\Sigma}_{\rm PE}$. Figure~\ref{fig:r_sigma} shows the evolution of the gas surface density around a low-mass star with a mass of $0.08\,M_\odot$. 
Owing to the effect of disk winds, the gradient of the surface density differs from the simple power-law distribution such as the minimum-mass solar nebula (MMSN).
We assumed that an inner disk edge is located at $r = 0.015\,{\rm au}$, which is the same as used by \citet{2019A&A...631A...7C}.
\citet{nakano21} found that the disk lifetime and its dependence on stellar mass agree with those obtained by observation \citep[e.g.,][]{2012A&A...547A..80B}. They also showed that an essential feature of the SED slope for transitional disks \citep[e.g.,][]{2018MNRAS.477.5191R} can be reproduced by their disk model.

\subsection{Initial condition}
We performed the \textit{N}-body simulations of planet formation for 3 Myr. The simulations started with planetary embryos. For the mass of initial planetary embryos, we simply adopted the isolation mass \citep{2002ApJ...581..666K},
\begin{equation}
\label{eq:miso}
M_{\rm iso} = 2 \pi a \Delta \Sigma_{\rm d}.
\end{equation}
The orbital separation between embryos was assumed to be $\Delta/r_{\rm H} =$7, 9, or 11, where $r_{\rm H}$ is the Hill radius.
The solid surface density is expressed by $\Sigma_{\rm d} \propto a^{q}$ and the slope was fixed to $q=-1.5$, unless otherwise stated.
If an expression for the Hill radius $r_{\rm H}\propto M_{\rm iso}^{1/3} a$ is assumed, Equation~(\ref{eq:miso}) can be written as $M_{\rm iso} \propto a^{3(q+2)/2}$. For $q=-1.5$, the isolation mass is proportional to $a^{3/4}$.
In this study, the initial total mass was usually set to $M_{\rm tot} = 6\,M_\oplus,$ comparable to the total mass of the planets found in the TRAPPIST-1 system.
The embryos were initially placed between 0.015 and 0.2\,au. This orbital region was located inside the water snowline at the initial time\footnote{In the disk evolution model we used, the disk formation from the molecular cloud core is considered \citep[e.g.,][]{2016MNRAS.461.2257K}. Therefore, the gas surface density in the inner region ($r \lesssim 1 {\rm \,au}$) is high ($\gtrsim 10^3 {\rm \,g\,cm^{-2}}$) in the early phase ($t_{\rm disk} \lesssim 10 {\rm \,kyr}$). In this phase, the snowline exists outside of 1\,au due to the viscous heating. As the disk evolves, the snowline moves inward from 1\,au.}; therefore, the embryos were assumed to be rocky, which agreed with the fact that the water mass fraction of the planets in the TRAPPIST-1 system would be smaller than 10wt\%.
Notably, it is not obvious that such initial conditions can be achieved, which should be investigated in a separate study (see the discussion in Section~\ref{sec:initial}).

Since our \textit{N}-body simulations start with planetary embryos, it is necessary to consider the time for their growth. Therefore, the time zero, $t = 0$, of the \textit{N}-body simulations is delayed by $t_0$ from $t_{\rm disk}=0$ in Figure~\ref{fig:r_sigma}. We typically used $t_0 = 10^3 {\rm \,yr}$, which is approximately $10^4$ orbits at $a=0.1 {\rm \,au}$ around a star with $M_* = 0.08 \,M_\odot$.

\subsection{Orbital migration}
As mentioned in Section~\ref{sec:intro}, the orbital migration significantly affects the final mass distribution of planetary systems. The type I migration torque is expressed as follows:
\begin{equation}
\Gamma = f_{\rm I} \left(\frac{M}{M_*}\right)^{2}
\left(\frac{H}{r}\right)^{-2} \Sigma_{\rm g} r^4 \Omega^2.
\end{equation}
For the formulae of disk-planet interactions, see \citet[][]{2017MNRAS.470.1750I,2018A&A...615A..63O,2019A&A...632A...7L}. The migration factor, $f_{\rm I}$, which determines the speed and direction of orbital migration, depends on the gradient of the local gas surface density, $p = d\ln \Sigma_{\rm g}/d\ln r$ \citep[e.g.,][]{2011MNRAS.410..293P}. 
When the slope is negative like the MMSN ($p=-3/2$), planets rapidly migrate inward. When the slope is more shallow (e.g., $p \simeq 0$), type I migration can be efficiently suppressed \citep[e.g.,][]{2015A&A...584L...1O}. Figure~\ref{fig:r_sigma} shows that the gas surface density exhibits a power-law-like distribution inside $r \simeq 0.1$ au in the initial phase ($t_{\rm disk} \lesssim 10^4 {\rm \,au}$). Afterward, the gas surface density slope is shallower than -3/2; thus, it is expected that planets undergo slower migration than was seen in previous studies \citep[e.g.,][]{2019A&A...631A...7C,2019A&A...627A.149S}.

\section{Results}\label{sec:results}
\subsection{Main result}\label{sec:main_result}

\begin{figure}
\centering
\includegraphics[width=\hsize]{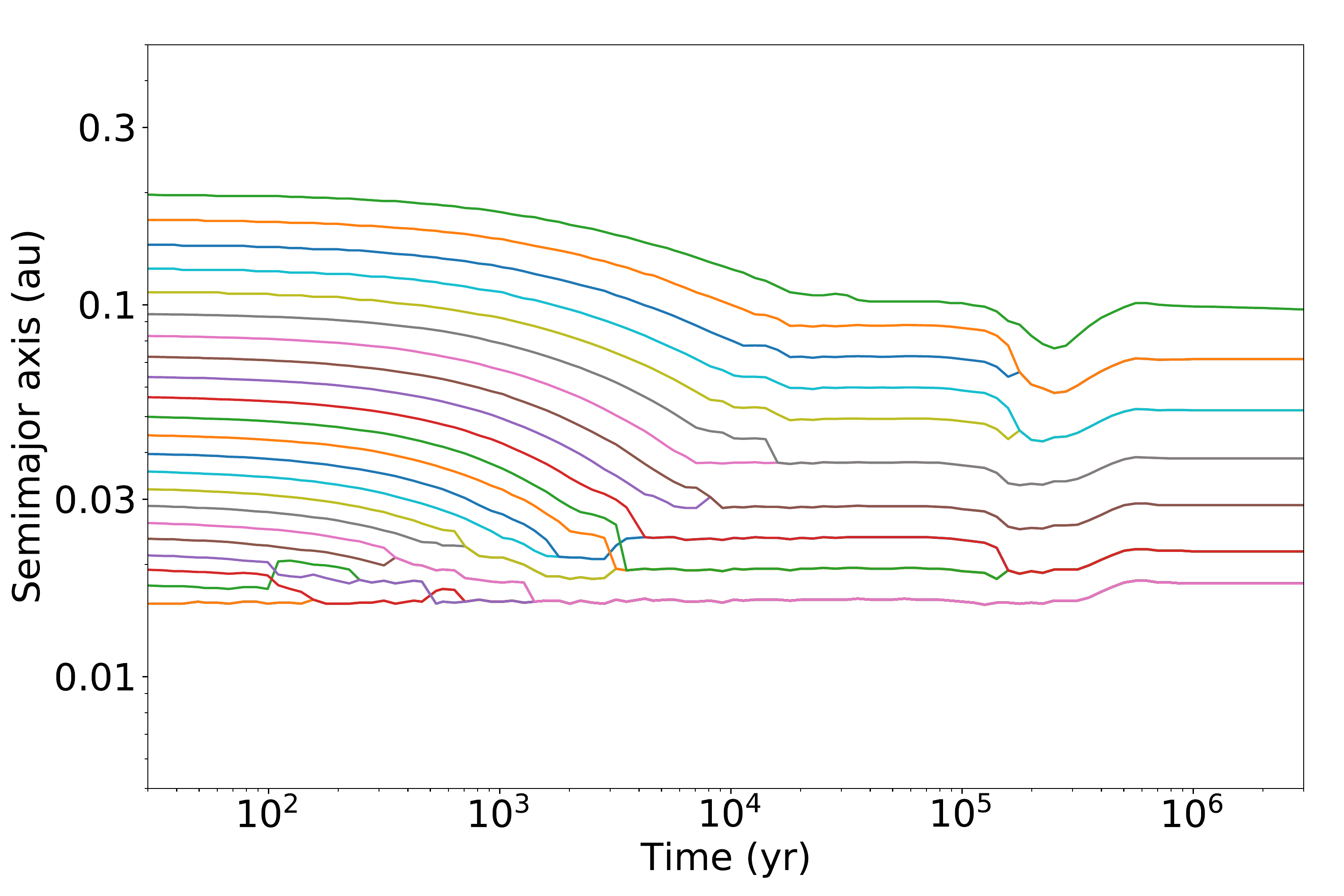}
\caption{Evolution of the semimajor axis for our main result. In this simulation, the initial orbital separation was set to $\Delta=7 \,r_{\rm H}$. The inner embryos underwent rapid migration, which enhanced the collision near the inner disk edge. Outer embryos exhibited slower migration and experienced fewer collisions.
}
\label{fig:fiducial}
\end{figure}

\begin{figure}
\centering
\includegraphics[width=\hsize]{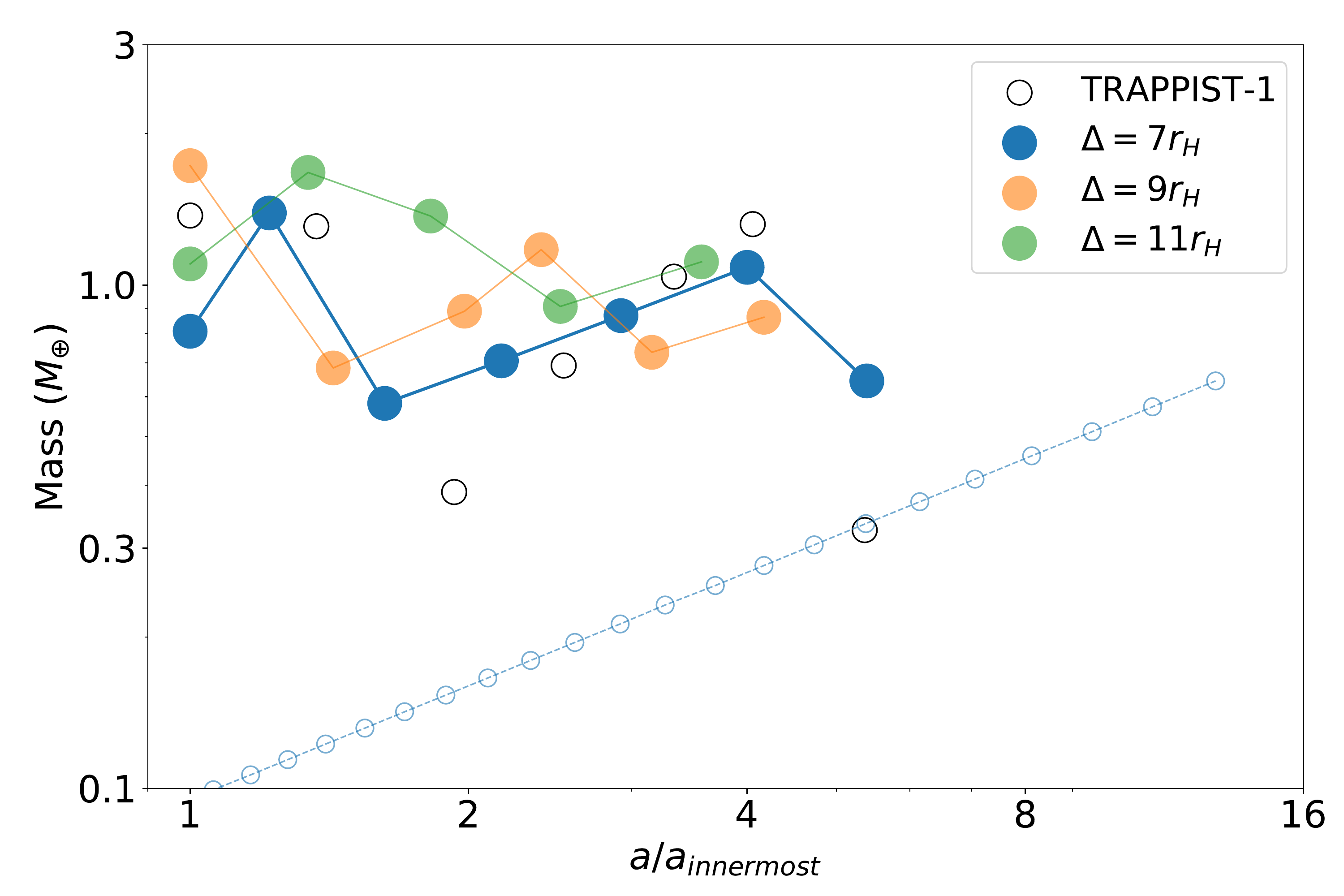}
\caption{Comparison of the final mass distribution. The result shown in Figure~\ref{fig:fiducial} is indicated by the blue symbols. The orange and green symbols are the results of simulations starting from $\Delta/r_{\rm H}=9$ and 11, respectively. The open circles represent the TRAPPIST-1 planets.
The small blue open circles represent the initial mass for $\Delta = 7 \,r_{\rm H}$.
}
\label{fig:fiducial2}
\end{figure}

First, we present the main result. Figure~\ref{fig:fiducial} shows the temporal evolution of the semimajor axis. The simulation was started with 23 embryos with the isolation mass (small blue open circles in Figure~\ref{fig:fiducial2}). For the initial orbital separation, we used $\Delta = 7 \,r_{\rm H}$.
Embryos initially in inner orbits ($a \lesssim 0.1{\rm \,au}$) underwent rapid inward migration on a timescale of approximately $10^3-10^4 {\rm \,yr}$, whereas embryos initially in outer orbits ($a \gtrsim 0.1 {\rm \,au}$) experienced slower migration on a timescale of approximately $\gtrsim 10^4 {\rm \,yr}$. During the rapid migration of inner embryos, collisions between the embryos were triggered near the inner disk edge, resulting in the growth of planets to large sizes. Moreover, outer embryos with slower migration did not frequently cause collisions as inner embryos did. After $t \sim 10^4 {\rm \,yr}$, planets were captured in mean-motion resonances with neighboring planets and their migration ceased. The figure shows is a slight outward migration at $t \simeq 10^5-10^6 {\rm \,yr}$ due to the positive slope of the gas surface density at $r \lesssim 1 {\rm \,au}$ (see also the discussion in Section~\ref{sec:resonance}).

Figure~\ref{fig:fiducial2} shows the mass distribution at the end of the simulation. The blue symbols show the result of the simulation shown in Figure~\ref{fig:fiducial}. We found that the two inner planets had relatively high masses, and the outer planets indicated reversed mass ranking in which the mass increases with orbital radius. The mass of each planet in the TRAPPIST-1 system is indicated by an open black circle. Comparing the mass distributions, we found that we could reproduce the two features of the mass distribution of the TRAPPIST-1 system: the inner planets are large and the outer planets show reversed mass ranking. Other colored symbols are examples of results of simulations with different initial orbital separations (i.e., $\Delta/r_{\rm H} =$ 9 or 11). These results also showed a tendency for the inside to be large and the outside to have an inverse mass ranking. Therefore, these results indicate that the mass distribution of the TRAPPIST-1 system could be reproduced by our formation model.

\subsection{Simulations with different parameters}

\begin{figure*}
\centering
\includegraphics[width=\hsize]{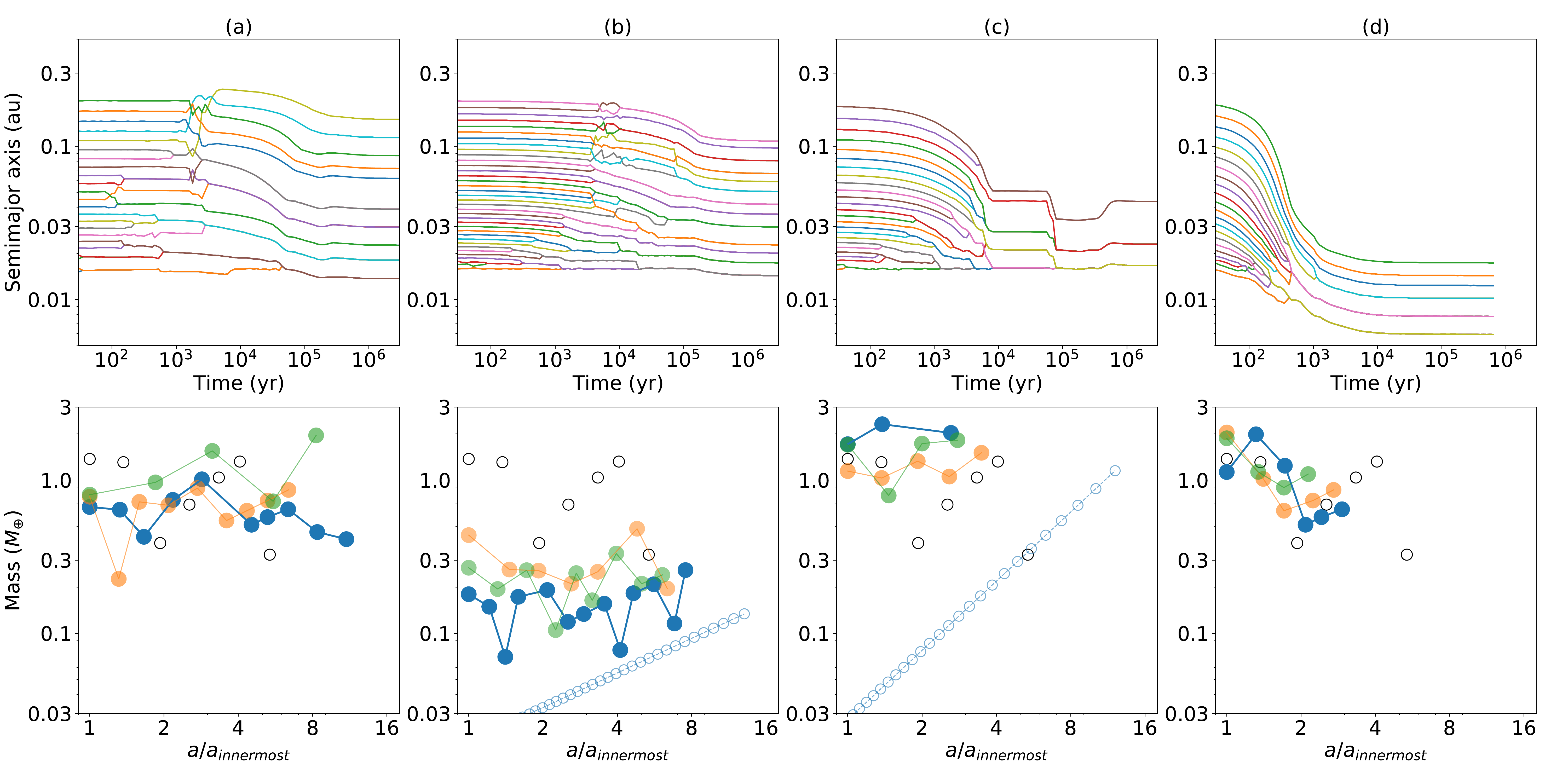}
\caption{Temporal evolution of the semimajor axis for simulations with initial separation of $\Delta = 7 \,r_{\rm H}$ as in Figure~\ref{fig:fiducial} (top).
Final mass distribution for simulations with $\Delta/r_{\rm H}=$7 (blue), 9 (orange), and 11 (green) as in Figure~\ref{fig:fiducial2} (bottom).
In each simulation, we (a) delayed the start time, (b) reduced the initial total mass, (c) changed the index of the initial embryo distribution, (d) and used the disk evolution with lower turbulent viscosity.
}
\label{fig:parameters}
\end{figure*}

We also performed simulations with different parameters to investigate whether the final mass distribution could change. First, we considered a delayed start time of the \textit{N}-body simulation. Figure~\ref{fig:parameters}(a) shows the result of a simulation with $t_0=10^5{\rm \,yr}$. In this case, we started the simulation from a state in which the gas surface density had already decreased to some extent (see $t_{\rm disk}=0.1{\rm \,Myr}$ in Figure~\ref{fig:r_sigma}), especially in the close-in region ($r \lesssim 0.1 {\rm \,au}$). In Figure~\ref{fig:parameters}(a), embryos initially in inner orbits also exhibit slow migration instead of rapid migration. As a result, collision and coalescence of inner embryos were not strongly triggered as in the case shown in Figure~\ref{fig:fiducial}.
In the bottom panel of the mass distribution, the reversed mass ranking is still visible, but the tendency for inner planets to become larger is less apparent. Therefore, it was likely that the rapid migration phase would be necessary to reproduce the mass distribution of the TRAPPIST-1 system.

Next, we performed simulations in which we varied the initial total mass of embryos. Figure~\ref{fig:parameters}(b) shows the result for $M_{\rm tot}=2\,M_\oplus$ , in which the total mass is reduced by a factor of three. Similar to the result of panel (a), the rapid migration of inner embryos was not profound as in Figure~\ref{fig:fiducial} because the mass of the initial embryos was lower by a factor of about five (see the small open blue circles in Figure~\ref{fig:parameters}(b) for the initial distribution with $\Delta = 7 \,r_{\rm H}$), and the migration of the embryos slowed down by that factor. For the final mass distribution, although it was not as pronounced as in panel (a), it was unlikely to be a mass distribution in which the inner planets became significantly larger. This result again suggested that rapid migration in the initial phase would be required to produce the characteristics of the mass distribution (inner large planets) of the TRAPPIST-1 system.
Notably, although not shown here, simulations were also performed for $M_{\rm tot}=4\,M_\oplus$. In this case, rapid and slow migration was observed in some simulations.

In addition, we performed simulations in which we changed the slope of the initial solid surface density. Figure~\ref{fig:parameters}(c) shows the result for simulations in which the slope of the initial solid surface density was set to $q=-1$. In this case, even outer embryos underwent rapid inward migration because when the initial slope was changed while fixing the total mass, the mass of outer embryos became higher than that for $q = -3/2$ (see the small open blue circles in Figure~\ref{fig:parameters}(c) for the initial distribution with $\Delta = 7 \,r_{\rm H}$). The outermost embryo was about two factors larger than for $q = -3/2$, leading to rapid migration. For the mass distribution, it did not show the reversed mass ranking due to rapid migration, which is inconsistent with the mass distribution of the TRAPPIST-1 system, suggesting that the slow migration phase is needed in addition to the rapid migration.
Although not shown, we also simulated a start time of the \textit{N}-body simulation at $t_0 = 10^5 {\rm \,yr}$ to the slow down the migration of the outer embryos. In this case, the rapid migration of outer embryos could be suppressed. However, collisions of inner embryos were also suppressed, and the inner planets did not grow enough. We  could not obtain a mass distribution with large inner planets, which agreed with the distribution of the TRAPPIST-1 system.

Finally, we present the results of simulations for a case with a different disk profile.
The turbulent viscosity was reduced by a factor of 100 (i.e., $\alpha_{r,\phi}=8 \times 10^{-5}$). The disk evolution is shown by the dashed curves in Figure~\ref{fig:r_sigma}.
In this case, the disk profile in the close-in region ($r \lesssim 1 {\rm \,au}$) was not determined by the viscous accretion, but by wind-driven accretion. As a result, the surface density slope was more positive in the close-in region. Examples of simulations are shown in Figure~\ref{fig:parameters}(d). Here, we used $t_0 = 2 \times 10^3 {\rm \,yr}$ for the start time of the simulations.
Embryos in the initial inner region ($a \lesssim 0.1 {\rm \,au}$) underwent very rapid inward migration, whereas outer embryos migrated more slowly than the inner embryos. The migration speed was faster than the speed shown in Figure~\ref{fig:fiducial} because the gas surface density was higher in the initial phase. In addition, inner embryos collided with neighboring embryos, leading to the growth of larger planets. For the final mass distribution, the inner planets were larger, and the outer planets exhibited reversed mass ranking, which was similar to the TRAPPIST-1 system.

Notably, in Figure~\ref{fig:parameters}(d), planets were not trapped at the inner disk edge at $r=0.015 {\rm \,au}$. As a result, the distribution of the semimajor axis at the end was not comparable to that of the TRAPPIST-1 system. The reason the planets went through the disk edge was the saturation of the corotation torque. The migration of planets could be halted at a steep surface density transition due to the positive corotation torque \citep[e.g.,][]{2006ApJ...642..478M,2021A&A...648A..69A}. When the viscosity was lower, the corotation torque could be saturated, resulting in the weakening of the planet trap \citep[e.g.,][]{2012ARA&A..50..211K}.
In addition, it was likely that the region near the inner disk edge was not in the dead zone, and the disk had a high viscosity. Therefore, the planet trap near the edge would be efficient in reality.

In summary, a mass distribution similar to that of the TRAPPIST-1 system could be achieved with rapid and then slow migration, although the planet trap at the inner disk edge should also be effective to reproduce the distribution of the semimajor axis. 
In this study, we showed that rapid and then slow migration was naturally achieved with a disk model that evolves with disk winds. Disk winds are not always necessary for this migration. Even in other disk models, it would be possible to explain the mass distribution of the TRAPPIST-1 system if a migration like this can be achieved, given that embryos are formed with a mass distribution that increases with orbital distance.

\section{Discussion}\label{sec:discussion}
\subsection{Mean-motion resonances}\label{sec:resonance}
\begin{figure*}
\centering
\includegraphics[width=0.8\hsize]{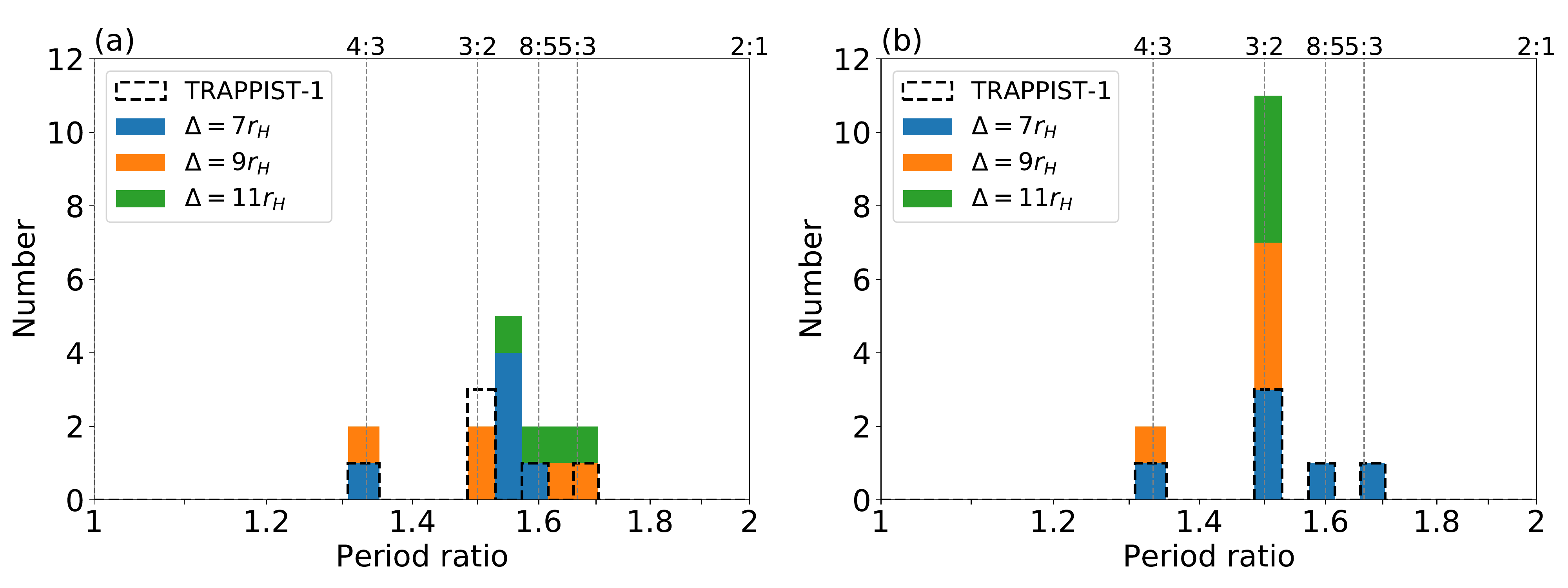}
\caption{(a) Histogram of the final period ratio of adjacent planets for simulations shown in Figure~\ref{fig:fiducial2}. The color scheme is the same as Figure~\ref{fig:fiducial2}. (b) Same as in panel~(a) with $t = 0.2 {\rm \,Myr}$.}
\label{fig:pratio}
\end{figure*}

The TRAPPIST-1 system is characterized by resonant relations. Previous studies \citep[e.g.,][]{2019A&A...631A...7C,2021ApJ...907...81L,2021arXiv210910984H} investigated how the resonant relations are reproduced. Although this study does not focus on reproducing the resonance relations, we also investigated whether the period ratio agrees with that of the TRAPPIST-1 system.
Figure~\ref{fig:pratio}(a) shows the final period ratio of adjacent planets for the three simulations shown in Figure~\ref{fig:fiducial2}. As stated in Section~\ref{sec:intro}, the orbital period ratios of neighboring planets in the TRAPPIST-1 system are in near first-order resonances (4:3 or 3:2) and higher-order resonances (8:5 or 5:3). The results of our simulations also showed that the orbital period ratio was between 4:3 and 5:3, which was approximately consistent with that of the TRAPPIST-1 system.

Although the period ratio approximately matched that in the TRAPPIST-1 system, some planets were not in the exact resonances at the end of the simulations. We briefly discuss this below.
In our main result shown in Section~\ref{sec:main_result}, the planets were captured in mean-motion resonances (typically 4:3 or 3:2) due to orbital migration at $t \sim 10^4 {\rm \,yr}$. For planets to be captured in mean-motion resonances, orbital migration is required. If the migration speed is too fast, the planets cannot be captured in resonances. According to previous studies, it is estimated that the critical relative migration timescale between embryos with the masses of $0.1-1 \,M_\oplus$ at $a \simeq 0.03 {\rm \,au}$ for the capture into 3:2 or 4:3 resonances in the TRAPPIST-1 system is approximately $10^2-10^3 {\rm \,yr}$ \citep[][]{2013ApJ...775...34O}.
In our main result shown in Figure~\ref{fig:fiducial}, the relative migration timescale at $t \sim 10^4 {\rm \,yr}$ was approximately $> 10^4 {\rm \,yr}$, which was sufficiently longer than the critical migration timescale for the resonant capture.
Figure~\ref{fig:pratio}(b) shows the period ratio of adjacent planets at $t = 0.2 {\rm \,Myr}$. At this stage, it agrees well with the resonant relations in the TRAPPIST-1 system. (The order of the resonant relations is not the same as in the TRAPPIST-1 system.)
The fact that the TRAPPIST-1 planets are captured in 4:3 and 3:2 resonances suggests that the orbits of the outer planets were not determined in the early phase when the relative migration speed is faster than the critical migration speed (i.e., when the disk gas surface density is high).

However, planets tended to depart from the exact resonance during the outward migration seen at $t \simeq 10^5-10^6 {\rm \,yr}$. Figure~\ref{fig:r_sigma} shows that the slope of the surface density in the close-in region became positive at $t_{\rm disk} \simeq 10^5-10^6 {\rm \,yr}$, leading to outward migration. At this stage, outer planets tended to be massive, hence when outer planets are more massive, they undergo faster outward migration. As a result, the planets experienced divergent migration and moved out of the exact resonance.
Notably, the disk evolution used in the simulation for the main result was derived for a specific parameter set (one of the typical sets in \citealt{2016A&A...596A..74S}), and the evolution of disk profiles could change depending on the parameter setting. 
Therefore, by varying the parameters of the disk evolution, it was possible to obtain a disk evolution in which no outward migration was observed. Although not shown here, we performed additional simulations in which the surface density slope was close to flat, but not positive. In the results of the simulations, we observed rapid and then slow migration, but no outward migration. As a result, the planets remained in exact mean-motion resonances at the end of the simulations, and the final period ratio distribution was similar to that shown in Figure~\ref{fig:pratio}(b).
Thus, both the mass distribution and resonant relation could be reproduced depending on the disk evolution.

Finally, we also comment on the Laplace resonances found in the TRAPPIST-1 system. It has been pointed out that the seven planets in the TRAPPIST-1 system are also in three-body resonances \citep{2017NatAs...1E.129L}. The Laplace resonance is characterized by the libration of the resonant angle ($\phi = p \lambda_1 - (p+q) \lambda_2 + q \lambda_3,$ where $\lambda$ indicates the mean longitude). In the simulation shown in Figures~\ref{fig:fiducial} and \ref{fig:pratio}, the libration of the same resonant angles $(p, q) = (2, 3)$ and $(1, 1)$ was observed. However, the angles tended to circulate during the divergent migration in the late phase. The origin of the Laplace resonance in the TRAPPIST-1 system has been studied \citep[e.g.][]{2018MNRAS.476.5032P,2021arXiv210910984H}, and it is likely that additional dissipation mechanisms are required to reproduce all resonant angles. Future studies of the evolution of three-body resonances by combining simulations of the formation stage and evolution stage are needed.

\subsection{Caveat for the initial condition}\label{sec:initial}
We considered embryos with an isolation mass (Eq.~\ref{eq:miso}) as an initial condition. This initial condition of increasing mass toward the outer orbit helped explain the reversed mass ranking of the TRAPPIST-1 system. However, it was unclear whether embryos were actually formed in this state. 

According to previous simulations of population synthesis, embryos grow to the isolation mass \citep[][]{2021arXiv210504596B}. We also performed test simulations that started with smaller initial embryos. The results showed that we could obtain a result similar to the main result, even when the initial mass was changed. These simulations might justify our assumption for the initial condition to some extent.
However, for the population synthesis calculations, the growth of embryos was calculated based on a simple semi-analytical model. In our test calculations, the initial mass was reduced by only a few factors, which did not allow us to follow the growth phase in detail. Although there are several previous studies of \textit{N}-body simulations of planet formation around low-mass stars \citep[e.g.,][]{2007ApJ...669..606R,2009ApJ...699..824O,2015ApJ...804....9C}, there are no studies that follow the growth from planetesimals in detail. Therefore, this process is not well understood.
Since this study aims to show how the mass distribution of the TRAPPIST-1 system can be explained under this assumption, we did not examine whether this initial distribution can be achieved.
Under which conditions this initial distribution is actually achieved needs to be investigated in a future publication.

\section{Conclusions}\label{sec:conc}

The mass distribution of the planets in the TRAPPIST-1 system has two characteristics: the inner two planets are large, and the outer planets show reversed mass ranking. Previous formation models of the TRAPPIST-1 system cannot explain these features in the mass distribution. Therefore, we investigated the formation of the TRAPPIST-1 system with \textit{N}-body simulations using a disk model that evolves with disk winds. The simulations were started from embryos with an isolation mass in which the outer orbits are more massive.
The results show that embryos in inner orbits undergo rapid migration, which accelerates the collision between neighboring planets near the inner edge of the disk, leading to the growth of large planets. 
Meanwhile, embryos in outer orbits experience slower migration and are more likely to be captured in mean-motion resonance with inner planets than to collide with them. Thus, the initial mass distribution characterized by the reversed mass ranking is more easily maintained. 
As a result of this rapid-then-slow migration, two features of the mass distribution of the TRAPPIST-1 system can be reproduced.
The rapid-then-slow migration is naturally achievable in the disk wind model. It is necessary to investigate whether this migration is also feasible with other disk models \citep[e.g.,][]{2017A&A...604A...1O,2019A&A...631A...7C,2021arXiv210504596B}.

We performed additional simulations with different parameter settings. Then, we find that it is difficult to reproduce the two features of the mass distribution in the TRAPPIST-1 system when the migration is constantly rapid or slow during formation. It is also suggested that the planet trap near the disk inner edge may be necessary.
We assumed an initial distribution of embryos with an isolation mass, which helps to explain the reversed mass ranking. Further studies are needed to determine whether the initial distribution with this reversed mass ranking can be achieved.

\begin{acknowledgements}
We thank the anonymous referee for helpful comments.
Numerical computations were in part carried out on PC cluster at Center for Computational Astrophysics of the National Astronomical Observatory of Japan.
This work was supported by JSPS KAKENHI grant Nos. 17H01105, 18K13608, 18H05438, 19H05087, and 21H00033.
\end{acknowledgements}

\bibliography{reference}
\bibliographystyle{aa}

\end{document}